\documentclass[twocolumn,aps,amsmath,amssymb,showpacs,pre,superscriptaddress]
              {revtex4} 
\usepackage{graphicx}

\begin{document}
\title{Solution of an associating lattice gas model with density anomaly on 
  a Husimi lattice }
\author{Tiago J. Oliveira}
\altaffiliation[Present and permanent address:]{Departamento de
  F\'{\i}sica, Universidade 
  Federal de Vi\c{c}osa, 36570-000, Vi\c{c}osa, MG, Brazil} 
\email{tiago@ufv.br}
\affiliation{Instituto de F\'{\i}sica
}
\author{J\"urgen F. Stilck}
\email{jstilck@if.uff.br}
\affiliation{Instituto de F\'{\i}sica and
National Institute of Science and 
Technology for Complex Systems\\
Universidade Federal Fluminense\\
Av. Litor\^anea s/n\\
24210-346 - Niter\'oi, RJ\\
Brazil}
\author{Marco Aur\'elio A. Barbosa}
\affiliation{Faculdade UnB Planaltina\\
Universidade de Bras\'{\i}lia\\
73300-000 - Planaltina, DF\\
Brazil}

\date{\today}

\begin{abstract}
We study a model of a lattice gas with orientational degrees of
freedom which resemble the formation of hydrogen bonds between the
molecules. In this model, which is the simplified version of the
Henriques-Barbosa model, no distinction is made between donors and
acceptors in the bonding arms. We solve the model in the
grand-canonical ensemble on a Husimi lattice built with hexagonal
plaquettes with a central site. The ground-state of the model, which
was originally defined on the triangular lattice, is exactly
reproduced by the solution on this Husimi lattice. In the phase
diagram, one gas and two liquid (high density-HDL and low density-LDL)
phases are present. All phase transitions (GAS-LDL, GAS-HDL, and
LDL-HDL) are discontinuous, and the three phases coexist at a
triple point. A line of temperatures of maximum density (TMD) in the
isobars is found in the metastable GAS phase, as well as another line
of temperatures of minimum density (TmD) appears in the LDL phase, part of
it in the stable region and another in the metastable region of this
phase. These findings are at variance with simulational results for the
same model on the triangular lattice, which suggested a phase diagram
with two critical points. However, our results show very good
quantitative agreement with the simulations, both for the
coexistence loci and the densities of particles and of hydrogen
bonds. We discuss the comparison of the simulations with our
results.  
\end{abstract}

\pacs{05.50.+q,61.20.Gy,65.20.-w}

\maketitle

\section{Introduction}
\label{intro}
The introduction of orientational degrees of freedom in lattice gas
models may result in rich phase diagrams. As an example, we may
mention the study of lattice gas models with direction dependent
interactions which were found to exhibit closed loop coexistence
curves \cite{aw79}, such as the ones found in solutions of glycerol
with guaiacol \cite{m23}, $m$-toluidine \cite{p24}, and
ethylbenzylamine \cite{m23}, which exhibit a nearly symmetric
coexistence loop with both 
an upper and a lower critical solution temperature. It was suggested
by Hirschfelder, Stevenson, and Eyring \cite{hse37}, that the
low-temperature critical point might be due to a highly directional
short-range interaction, such as a hydrogen bond: while at low
temperature the ordering of these interactions lowers the energy of
solution, with rising temperature this ordering is decreased and phase
separation occurs. These suggestion was followed in the model proposed
by Barker and Fock some time later \cite{bf53}, and the solution of
this model in the quasi-chemical approximation actually displays a
closed coexistence loop. A simplified version of the model defined on
a conveniently decorated simple cubic lattice, may be mapped on the
three-dimensional Ising model and thus much
precise information is known about its thermodynamic
behavior \cite{w75}. When the directionality of the part of the 
interactions in the model due to the hydrogen bonds is increased, the
results are closer to the 
experimental data for the mixtures cited above, although the
correspondence to the Ising model is lost \cite{aw79}. 

In water, the ordering of hydrogen bonds is supposed to be important
in determining the unusual thermodynamic and dynamic behavior, 
including the possible existence of an experimentally unaccessible
liquid-liquid phase transition \cite{d03}.  Liquid-liquid phase 
transitions were originally found by Monte Carlo simulations of
realistic liquid water models with atomic details \cite{p92,h97}, but
they were already observed experimentally 
in systems such as phosphorus \cite{k00}, triphenyl
phosphite \cite{k04} and $n$-butanol \cite{k05,t04}.  
Tetrahedral liquids, such as silica and water,  also present thermodynamic
and dynamic anomalies which can possibly be related to the second
critical point (SCP) associated with these 
transitions \cite{e01,s06,s02,x05,k07}. Among these anomalous
features, we note the 
increase of density with 
temperature that happens in liquid water at temperatures below $4^oC$
and the apparent divergent 
behavior of thermodynamic response functions with decreasing
temperatures towards 
the deep super-cooled liquid, at atmospheric pressures \cite{d03}.

Several lattice models with orientational interactions, usually called
network-forming fluids or associating lattice gases, have been proposed
in two- \cite{b70,b04,h05,s93,gs02} 
and three-dimensions \cite{b72,b97,r96,p04,g07} to investigate the
thermodynamic anomalies presented by water and tetrahedral liquids.
Some of them were also found to present dynamic anomalies similar to
the ones in 
liquid water \cite{s07,gs07,kf08}. In these models, the
distinction between hydrogen bonds and van der Waals interactions is
essential: these two main ingredients contribute to
the appearance of a competition between distinct molecular states
presenting high density (lowly bonded) and low density (highly bonded) 
structures.  
Nevertheless, in many models, more specific molecular interactions,
and structures, are used to bias the system towards
a low density liquid (LDL), at low pressures, or a high density
liquid (HDL), at high pressures.  
As an example, some models use many-body interactions to unfavor
molecular packing in the neighborhood of a hydrogen
bond \cite{b04,b72,r96,p04}.  
Others actually energetically favor the LDL states through a repulsive
van der Waals interaction \cite{b97,h05,g07}.  
Some implement fluctuating bonding structures \cite{gs02}, 
additional unbounded molecular states (to stabilize a disordered
anomalous liquid) \cite{b04,r96,p04}  
and some even use \textit{ad hoc} variations of volume with bond
formation \cite{gs02}. 

Considering the increasing complexity found in models for water
in the literature \cite{hb},   
simple three- and two-dimensional models of liquid water, including
only van der Waals and hydrogen bond interactions, have been
investigated with the aim of 
finding the minimal requirements for water-like anomalous behavior
\cite{h05,g07,s07,gs07,hg05,b07,b08}. One of these models, the GBHB
model proposed by Girardi et al.\cite{g07}, is a three-dimensional
fluid, defined on a body centered cubic lattice, with first-neighbor
van der Waals and hydrogen-bond like
interactions. It presents a phase diagram with two distinct liquid phases
(high-density and low-density liquid - HDL and LDL) besides a GAS
phase. Two coexistence lines (GAS-LDL and LDL-HDL) ending at critical
points were originally found with Monte Carlo 
simulations by Girardi et al.\cite{g07}.  Also, the isobars present
temperatures 
of maximum density (TMD) on a line in the pressure-temperature plane,
resembling qualitatively the scenario emerging from the simulations by
Poole {\textit et al} \cite{p92}. Nevertheless, a qualitatively
different phase diagram was 
found for the same model in a recent work by Buzzano and  
collaborators \cite{bs08}, in which the phase diagrams of a
three-dimensional model of network forming 
fluid \cite{g07} were investigated using the cluster variational method
\cite{p03}. With this approach 
they were able to show that the topology of the phase diagram of the model
was much more complex than originally found with Monte Carlo
simulations but, at the same 
time, very diverse from  the one expected for water. 
It was found that the so-called critical points were indeed
tricritical points connected to a line of critical points. Besides
that, another line of critical points was found separating the GAS and
HDL phases, terminating in a critical end point on the GAS-LDL
coexistence curve.

In a more recent paper \cite{p09} from the same group, the
previous analysis was extended by including another two
three-dimensional 
models of `liquid water', also defined on the bcc lattice,
originally proposed by Bell \cite{b72} and by Besseling and
Lyklema \cite{b97}. They revisited the three models using the same
methodology and the same conclusion holds for them: in all cases the
phase diagrams were indeed much more complex than originally expected.
In the previous analytical studies \cite{b72,b97}, phase diagrams were
oversimplified due to a `homogeneity' assumption on the lattice
sites, and by allowing sublattice ordering, more stable ordered phases
appear and the disordered, homogeneous and water-like fluid becomes
either unstable or metastable \cite{p09}. 

Here we investigate a simplified version of a two dimensional
associating lattice gas model on the core of the Husimi
cactus \cite{h05},  
considering these recent results on lattice models with water-like behavior. 
The original model was proposed by Henriques and Barbosa and studied
through Monte Carlo simulations in a series of
papers \cite{h05,hg05,b07,s07,s09}. 
In the Henriques-Barbosa model each site of a triangular lattice can be
occupied by a water molecule or empty. A molecule has
four bonding arms (two donors and two acceptors) and two inert
arms separated by an angle of $180^o$. All arms lie on lattice edges.
A HDL was found at low temperatures and high pressures for 
repulsive van der Waals interactions, while a LDL was found at low
temperatures and lower pressures.  
The first Monte Carlo simulations provided 
indications of a coexistence between the HDL and the LDL, with the presence of
a second critical point (SCP) at the end of the HDL-LDL coexistence
locus \cite{h05,hg05}. A temperature of maximum density 
 was also found in the neighborhood of this 
SCP. Variations of this model were also investigated through Monte
Carlo simulations: the distinction between donors and acceptors was
excluded from the model 
and distortions were introduced in the bonding arms \cite{b07}. In all
cases, the SCP and a line 
of TMD were found to be present, in an indication of the apparent
robustness of these features 
in the phase diagram. More recently, the phase diagram of the
Henriques-Barbosa model was revisited using simulations 
and it was found to be much more complex and richer than originally
observed \cite{s09}. 
The new simulations suggest that the GAS-LDL coexistence curve ends at a
tricritical point, and that 
the LDL-HDL coexistence ends at a bicritical point, where the two
continuous transition lines (GAS-LDL and GAS-HDL) also
meet \cite{s09}.  

In this work we consider the version of the Henriques-Barbosa model
without distinction between proton donors and acceptors \cite{b07}. 
This simplifying assumption does not lead to essential differences in
the phase diagrams of this model, particularly with  
respect to the presence of the density anomaly and the HDL-LDL first order
phase transition \cite{b07}. The  Husimi cactus is built with  
hexagonal plaquettes with a central site (composed by six elementary
triangles) as base cells, hereafter called hexagons only. This may be
seen as a second-order approximation on the triangular 
lattice \cite{g95}. Hexagons were chosen  as a base cells because
they are the simplest alternative we found to 
reproduce exactly the ground state of both ordered phases (LDL and
HDL) on the triangular lattice.  
We advance that the phase diagram of the Henriques-Barbosa model we obtained
turned out to be very different from the one originally
obtained with Monte Carlo 
simulations \cite{b07}. Nevertheless, it is closer to the more recent
simulations of 
the model with distinction between donor and acceptor arms \cite{s09}. In  
our study, the GAS-LDL and LDL-HDL coexistence lines developed into two
first order phase transitions ending at a triple point. In addition 
to this, a novel first order transition line between the GAS and HDL
phases appeared separating both phases for all pressures. 
Although our results show that the Henriques-Barbosa model may have a
complex and intriguing 
phase diagram, in the current formulation the model seems to be
inappropriate for liquid water. Nevertheless it does present some
water-like features such as a temperature of maximum density
  (TMD) in the fluid phase,  
which can be used as a starting point for more
complex two-dimensional models of liquid water. 

In our opinion, in
models for complex fluids the combination of approximate calculations
with extensive numerical simulations are complementary in the study of
their thermodynamic behavior. Although approximate analytical results
may be at variance with
the correct ones for the corresponding model, they may also suggest
more detailed numerical studies of the model to ascertain that the
real behavior is found. Besides, it is remarkable that a very good
agreement was found between the simulational and the
cluster-variational results for the 3D associating lattice gas in
\cite{bs08}. As will be shown later, this is also true for the
2D model studied here.

This paper is organized as follows. In section \ref{mod} the model is
introduced in more detail on the triangular lattice and its ground
state is analyzed.  
We then proceed defining
the model in a Husimi lattice built with hexagons, such that the ground
state properties on the triangular lattice are exactly reproduced on
the Husimi lattice. We also present the solution of the model in terms
of recursion relations and the calculations of the grand-canonical
potential in the bulk of the tree. In section \ref{tp} the
thermodynamic properties of the model are studied and compared with
Monte Carlo simulation data found in the literature for the same model. 
Final discussions and the conclusions may be found
in section \ref{conclu}.

\section{Definition of the model and solution on the Husimi lattice}
\label{mod}
We consider the simplified version of the Henriques-Barbosa model on
the triangular lattice. Each site 
of the lattice may be either empty or occupied by a single molecule. A
molecule has four bonding arms, without distinction between donors or
acceptors of protons, and two neutral (non-bonding) arms. The neutral
arms form an angle $180^o$, and therefore each 
particle has three possible orientations of the bonding
arms. Thus, we are considering the symmetric undistorted case
discussed in \cite{b07}. The possible configurations of a site
$i$ will be represented by a variable $\eta_i$, which vanishes if the 
site is empty and assumes the values 1, 2, or 3 if the site is occupied
in one of the possible orientations of the bonding 
arms. Repulsive van der Waals interactions 
$\epsilon>0$ exist between particles on first neighbor sites,
and an energy $\gamma<0$ corresponds to 
each hydrogen bond on the lattice. Therefore, if $|\gamma|>\epsilon$,
a pair of 
particles on first neighbor sites with an hydrogen bond between them
is associated to a net {\em negative} energy and thus the interaction
becomes attractive. Since we will study the model in the
grand-canonical ensemble, an activity $z=\exp(\mu/k_BT)$ corresponds
to each particle on the lattice, where $\mu$ is the chemical
potential. We may relate the parameters used here and those chosen in
reference 
\cite{b07}, there a pair of first-neighbor sites occupied by
particles with a 
hydrogen bond between them corresponds to an energy $-v$ and if no
hydrogen bond is present this energy is $-v+2u$
\cite{footn01}. Therefore, we have
$\epsilon=-v+2u$ and $\gamma=-2u$, and we
notice that for $u/v=1$ we have $|\gamma|/\epsilon=2$. Since the
simulations in \cite{b07} were done for this particular choice, we
restrict our numerical calculations to this particular case.

\begin{figure}
\begin{center}
\includegraphics[scale=0.6]{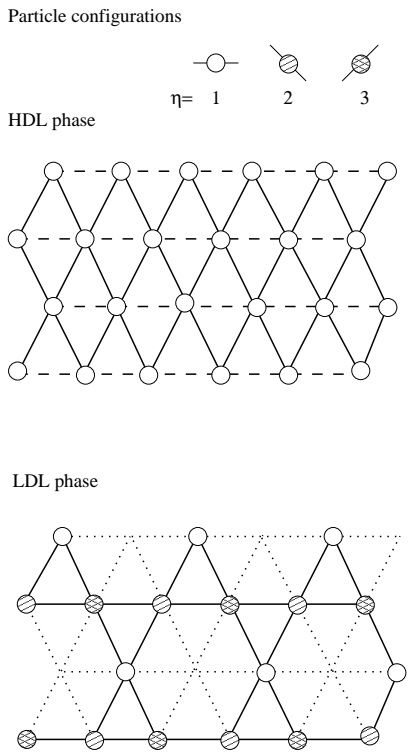}
\caption{A representation of the two ordered phases of the
    model with the configuration of each particle identified by the
    orientation of the inert arms (upper panel).  
    In both high density liquid (middle panel) and low density liquid
    phases (bottom panel) hydrogen bonds are indicated using full lines
    lattice edges with van der Waals interactions only are drawn with
    dashed lines, while the other lattice edges are represented by dotted
    lines.}  
\label{f1}
\end{center}
\end{figure}

Three phases were 
found in the ground state in earlier investigations \cite{h05,b07}:
The GAS phase corresponds to the empty 
lattice, and is stable at low values of the chemical potential; as the
chemical potential is increased, the low-density liquid (LDL) becomes
stable, in which a fraction $\rho=3/4$ of the sites are occupied by
particles and 
all lattice edges between two particles are occupied by hydrogen
bonds. For still higher chemical potentials, a high-density liquid
(HDL) becomes stable, in which all sites are occupied and therefore
$\rho=1$. In Fig. \ref{f1} both liquid phases in the ground state
are depicted.

\begin{figure*}
\begin{center}
\includegraphics[scale=0.6]{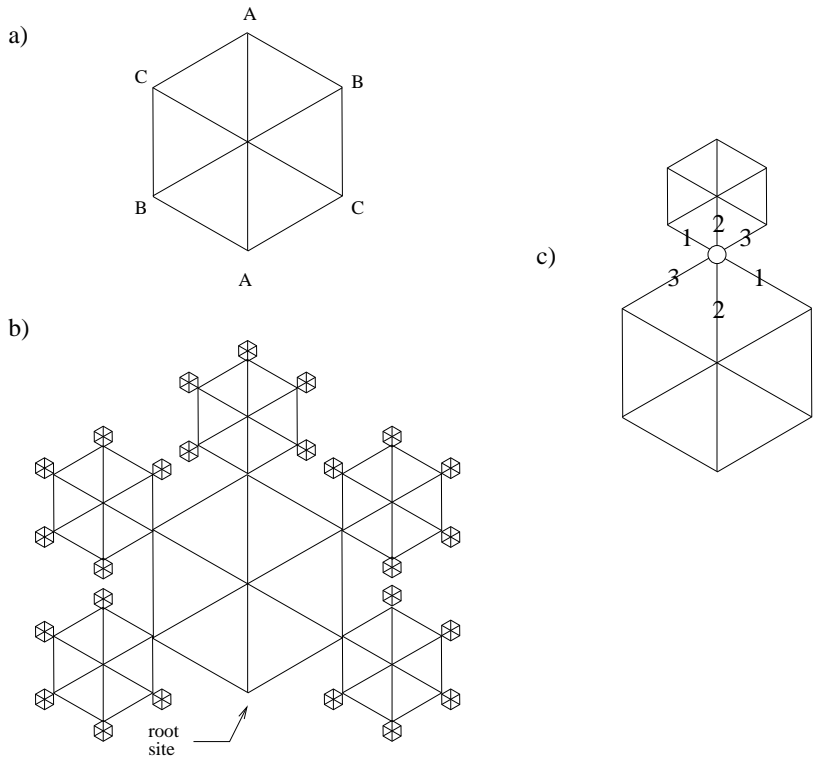}
\caption{a) Definition of the sublattices. b) A subtree with three
  generations. c) Definition of $\eta$ for a site occupied by a
  particle. The numbers on the lattice edges correspond to the values
  of the variable $\eta$ if the inert bonds of the particle are placed
  on these edges.} 
\label{f2}
\end{center}
\end{figure*}

To describe the
sublattice structure of the LDL phase it is necessary to  
introduce four sublattices: three of them formed by sites on the
borders of the hexagons and the central sites (see
  Fig.~\ref{f2}~(a)). This 
immediately leads to a fourfold  
degeneracy (corresponding to the placement of a hole in four different
sites), which is captured by the Husimi tree if the homogeneity assumption is
avoided~(\cite{p09}). The sublattices are equivalent in the HDL but we
take care of the orientational order. This leads to a
threefold degeneracy (corresponding to possible orientations for water
molecules in the lattice), and this behavior is  also captured by the
tree without the homogeneity assumption. As will be shown later, the
states of the central site of each hexagonal plaquette will always be
summed in the recursive equations obtained from the hierarchical
structure of the lattice. Considering
this (and also to simplify our notation), we use only the three
sublattices for the sites on the perimeters of the hexagons, as is
shown in Fig. \ref{f2}.

Let us now
discuss the ground state of the model in some detail. In the GAS phase
all sites are empty and we will associate a vanishing energy to this
configuration, $E_{GAS}=0$. The LDL phase on the Husimi lattice is
fourfold degenerate, characterized by 
empty sites either at the center of each hexagon or at the sites of
one of the three sublattices A, B, and C. Recalling that all
edges between first neighbor sites occupied by particles have hydrogen
bonds on them, the energy per hexagon (including the chemical
potential term) will be: 
\begin{equation}
E_{LDL}=6(\epsilon+\gamma)-3\mu,
\end{equation}
where we remember that each particle on the vertices of
the hexagons is shared by two plaquettes.
In the HDL phase, all sites are occupied and 8 of the 12 edges of each
hexagon are occupied by hydrogen bonds, while the remaining 4 are
not. Thus, there are three possible configurations of the hydrogen
bonds. The energy per hexagon in this phase is:  
\begin{equation}
E_{HDL}=12\epsilon+8\gamma-4\mu.
\end{equation}
It is easy to find which phase corresponds to the minimum energy for
given parameters $\epsilon$, $\gamma$, and $\mu$. Using the vdW interaction
$\epsilon$ as the energy scale, we may define the dimensionless
variables ${\bar \gamma}=|\gamma|/\epsilon$ and ${\bar
  \mu}=\mu/\epsilon$. The ground state corresponds to the GAS phase if
${\bar \mu}<2(1-{\bar \gamma})$, to the LDL phase if $2(1-{\bar
  \gamma})<{\bar \mu}<2(3-{\bar \gamma})$, 
and to the HDL if 
${\bar \mu}>2(3-{\bar \gamma})$. As observed above, these values are
the same as the ones 
found for the ground state on the triangular lattice \cite{b07}.

\subsection{Recurrence relations on the Husimi cactus}

As usual, we start defining partial partition functions for rooted
subtrees, fixing the configuration of the root. One of these subtrees
is shown in Fig. \ref{f2}. There are $3 \times 4$ configurations of
the root sites, so we define 12 partial partition functions $g_i$,
$i=1,2,\ldots,12$, where the index $i$ stands for the root site
configuration. We may associate the configurations $(s,\eta)$, where
$s=A,B,C$ stands for the sublattice and $\eta=0,1,2,3$ for the site
configuration, to the indices $i$ of the partial partition functions
following the convention indicated in table \ref{tabi}. It is also
useful to define the configurations of the bonding 
arms of a particle in a way which may be applied to all sites of the
tree (only sites at the perimeter of the hexagons are considered,
since we will sum over the configurations of the central sites). We
therefore consider a particular site with a particle on 
it, which will belong to two hexagons in different generations of the
tree, and imagine that we circle around the site clockwise, starting
{\em outside} the hexagons. The configuration variable $\eta$
associated to this site 
will be equal to the number of lattice edges we cross until
the one where one of the inert arms of the particle are
located is reached, added with one. This 
definition is illustrated in Fig. \ref{f2}. We proceed considering
the operation of attaching 5 subtrees with $M$ generations to a new
root hexagon, building a subtree with $M+1$ generations. Summing over the
$4^5=1024$ possible configurations of the root sites of the
$M$-generations subtree, we will arrive to recursion relations for the
partial partition functions, which are of the form:
\begin{equation}
g_i^{M+1}=\sum_{j=1}^{1024} \left(\sum_{k=0}^3 z^{n_{j,k}}
\omega_p^{p_{i,j,k}} \omega_b^{b_{i,j,k}}\right) \prod_{\ell=1}^{12}
(g_\ell^M)^{e_{i,j,\ell}}, 
\label{rr}
\end{equation}
where $n_{j,k}$, $p_{i,j,k}$, and $b_{i,j,k}$ are the number of
particles, 
number of pairs of particles in first neighbor sites and number of
hydrogen 
bonds for each contribution $j$ to the partial partition function of
$g_i^{M+1}$, given that the configuration $\eta$ of the central site is
equal to $k$. $\omega_p=\exp[-\epsilon/(k_BT)]$ and
$\omega_b=\exp[-\gamma/(k_BT)]$ are the Boltzmann factors associated
to the van der Waals interactions and hydrogen bonds, respectively.
We notice that the activity of the particle which 
eventually  is placed on the root site is not considered at this
level. The exponents $e_{i,j,\ell}$ assume integer values between 0 and
2. We remark that for each pair of indices $(i,j)$ there will be at
most five
nonzero exponents $e_{i,j,\ell}$, since this is the number of subtrees 
with $M$ generations linked to the new hexagon at the root. The
exponents $e_{i,j,\ell}$ depend of the index $i$ only because the
number of incident subtrees with root sites in each sublattices
depends of the sublattice of the root site of the new subtree.

\begin{table}
\begin{ruledtabular}
\begin{tabular}{ccc}
$(s,\eta)$&$i$& ${\bar i}$ \\
\hline
$(A,1)$, $(A,2)$, $(A,3)$&$1$, $2$, $3$& $10$\\
$(B,1)$, $(B,2)$, $(B,3)$&$4$, $5$, $6$& $11$\\
$(C,1)$, $(C,2)$, $(C,3)$&$7$, $8$, $9$& $12$\\
$(A,0)$, $(B,0)$, $(C,0)$&$10$, $11$, $12$& $-$ \\
\end{tabular}
\end{ruledtabular}
\caption{To each possible state of a site, specified by its
  sublattice $s$ and 
  configuration $\eta$ (first column), an index
  $i$ is assigned (second column).  The indexes ${\bar i}$
  which appear in the denominator of eq.~(\ref{rrr}) are shown on the third
  column.} 
\label{tabi}
\end{table}

In similar calculations, often the recursion relations are obtained by
hand, usually summing the contributions with some graphical aid. In
the present case, due to the large number of contributions, this 
procedure is very tedious and therefore errors are quite frequent.
Although we actually obtained the recursion relations
explicitly in this
way, using symmetries to generate the 
expressions for the recursion relations, these expressions are much too
large to be given here. To assure that the recursion relations are
free of errors, we also decided to write a rather simple code which
generates the sets of 24 integer numbers $n_{j,k}$, $p_{i,j,k}$,
$b_{i,j,k}$ and $e_{i,j,\ell}$ for each contribution $j$ to the recursion
relation for $g_i^{M+1}$, similar to what was done by Zara and Pretti
in a model for RNA on the Husimi lattice \cite{z06}. Since we are
interested in the behavior of 
the model in the thermodynamic limit, we should consider fixed points
of these recursion relations. As expected, however, the partial
partition functions diverge in this limit. So, we may define ratios of
these functions, which may approach a finite value as $M \to
\infty$. We thus define the ratios $R_i$, $i=1,2,\ldots,9$, dividing
each partial partition function with a particle at the root site by
the partial partition function with an empty root site in the same
sublattice. This leads us to the ratios $R_i=g_i/g_{\bar i}$, where
the values of ${\bar i}$ are shown in table \ref{tabi}. We may then
obtain recursion relations for the 
ratios from the ones for the partial partition functions,
Eq. \ref{rr}. They are: 
\begin{equation}
R_i^{M+1}=\frac{\sum_{j=1}^{1024} \left(\sum_{k=0}^3 z^{n_{j,k}}
\omega_p^{p_{i,j,k}} \omega_b^{b_{i,j,k}}\right) \prod_{\ell=1}^9
(R_\ell^M)^{e_{i,j,\ell}}}
{\sum_{j=1}^{1024} \left(\sum_{k=0}^3 z^{n_{j,k}}
\omega_p^{p_{{\bar i},j,k}} \omega_b^{b_{{\bar i},j,k}}\right) \prod_{\ell=1}^9
(R_\ell^M)^{e_{{\bar i},j,\ell}}}.
\label{rrr}
\end{equation}

\subsection{Densities in the core of the Husimi cactus}

In order to obtain densities in the central region of the
tree, we consider the operation of attaching 6 subtrees to the
central hexagon, which leads to an expression for the partition
function of the whole tree:
\begin{equation}
Y_M=\sum_{j=1}^{4096} \left(\sum_{k=0}^3 z^{N_{j,k}}
\omega_p^{P_{j,k}} \omega_b^{B_{j,k}}\right) \prod_{\ell=1}^{12}
(g_\ell^M)^{E_{j,\ell}},
\label{ym}
\end{equation}
where $N_{j,k}$, $P_{j,k}$ and $B_{j,k}$ are the total number of
particles, nearest neighbors and hydrogen bonds 
on the central hexagon with the configuration of the border sites
given by $j$ and the central site in the configuration $k$,
and $E_{j,\ell}$  
is the number of border sites with configuration $\ell$.

Again the set of integer exponents was generated by a computer
program, as well as manually, and both procedures lead to the same
final results. Now, for example, the density of particles (defined
here as the number of particles divided by the number of sites) in the
central hexagon will be given by:
\begin{equation}
\rho=\frac{z}{7Y_M}\frac{\partial Y_M}{\partial z},
\label{densi}
\end{equation}
where $\rho$ is in the range $[0,1]$. We notice that the activities of
the sites at the perimeter and the center of the central hexagon of
the tree are considered in expression (\ref{ym}), so the factor 7
assures the proper normalization of the density. In other words, the
numbers $N_{j,k}$ in expression (\ref{ym}) are in the range $[0,7]$. A
similar procedure leads to 
expressions for the densities of hydrogen bonds and van der Waals
interactions per site. Dividing both the numerator and the denominator
of the 
expressions for the densities by $(g_{10}^M\,g_{11}^M\,g_{12}^M)^2$ we
may express them in terms of the ratios and the parameters of the
model. Thus, for example:
\begin{equation}
\rho=\frac{1}{7}\frac{\sum_{j=1}^{4096} \left(\sum_{k=0}^3
  N_{j,k}z^{N_{j,k}} 
\omega_p^{P_{j,k}} \omega_b^{B_{j,k}}\right) \prod_{\ell=1}^{9}
(R_\ell^M)^{E_{j,\ell}}}{\sum_{j=1}^{4096} \left(\sum_{k=0}^3
z^{N_{j,k}} 
\omega_p^{P_{j,k}} \omega_b^{B_{j,k}}\right) \prod_{\ell=1}^{9}
(R_\ell^M)^{E_{j,\ell}}}.
\label{densi1}
\end{equation}

To obtain the thermodynamic behavior of the model, we may iterate the
recursion relations until a fixed point for the ratios $R_\ell$ is
reached with the required numerical precision, and then calculate the
densities at the center of the tree. The convergence of the recursion
relations generally is quite fast. In certain regions of the
parameter space, more than one fixed point may be stable, signaling
coexistence of phases. To locate the first order 
transition in such cases it is necessary to compare free energies of
different phases. An expression for the grand-canonical free energy is
obtained in what follows. 

\subsection{Grand-canonical free energy}

To obtain the grand-canonical free energy of the model in the core of
the tree, we may proceed following the prescription proposed by
Gujrati \cite{g95}. For this purpose, it is convenient to notice that
if we connect the central 
site of each hexagon to the central sites of the first-neighbor
hexagons, we end up with a Cayley tree with 
coordination $q=6$ and ramification $\sigma=q-1=5$. Now we may assume 
that the total free energy of the tree is the sum of the free energies
associated to each {\em hexagon}. This takes care of the sublattice
structure of the model. The hexagons may then be classified in
generations identified by the index $m$, starting with the ones placed
on the surface, for which $m=1$ and ending at the central site
($m=M$  for a tree with $M$ generations of hexagons). It is natural,
considering the structure of the tree, to 
assume a radial symmetry for the local free energies, so that we
will represent by $\phi_M(m)$ the free energy of a hexagon in the
$m$'th generation of a tree with a total of $M$ generations. The total
free energy of the tree will then be:
\begin{equation}
\Phi_M=\sum_{m=1}^M N_M(m)\phi_M(m),
\end{equation}
where $N_M(M)=1$ and
\begin{equation}
N_M(m)=q\sigma^{M-m-1},\;\;m=1,2,\ldots,M-1
\end{equation}
are the numbers of hexagons in the $m$'th generation. Now we may see
that:
\begin{eqnarray}
\Phi_{M+1}&-&\sigma \Phi_M=\phi_{M+1}(M+1)-\sigma \phi_M(M)+q \phi_{M+1}(M)  
\nonumber \\
&&+ q\sum_{m=1}^M \sigma^{M-m}[\phi_{M+1}(m)-\phi_M(m)].
\label{fepg}
\end{eqnarray}
It is now reasonable to assume that in the thermodynamic limit we
should have the free energies per hexagon in the core of 
the tree approaching a limiting bulk value $\phi_b$, so that
$\phi_{M+1}(M+1)=\phi_M(M)=\phi_{M+1}(M)=\phi_b$ for $M \to
\infty$. Close to the surface, the free energies per hexagon should be
functions of $m$, but we may assume $\phi_{M+1}(m)-\phi_M(m) \to 0$
when $M \to \infty$. Therefore, we may conclude from equation
(\ref{fepg}) that:
\begin{equation}
\phi_b=\frac{1}{2}(\Phi_{M+1}-\sigma \Phi_M),
\end{equation}
in the thermodynamic limit $M \to \infty$ and for the model we are
considering here $\sigma=5$. Actually, this expression may also be
obtained from a somewhat stronger assumption that the free energies
per hexagon in the thermodynamic limit should assume only two values:
$\phi_b$ on the surface $m=1$ and $\phi_b$ in all other cases
\cite{oss09}. Also, we notice that the original argument for the
calculation of the free energy was also recently generalized for
Husimi trees with a sublattice structure in Semerianov and Gujrati
\cite{sg05}. The derivation presented here is different from the ones
originally proposed in \cite{g95} and \cite{sg05}, but the results are the
same \cite{footn02}.

Since the partition
function on a $M$-generations tree is $Y_M$, we have:
\begin{equation}
\phi_b=-\frac{1}{2}k_BT\ln\left(\frac{Y_{M+1}}{Y_M^5}\right).
\label{phib}
\end{equation}
Substituting the partition function~(\ref{ym}) in this expression, and
expressing the sums in terms of the ratios, we obtain:
\begin{widetext}
\begin{equation}
\frac{Y_{M+1}}{Y_M^5}=\frac{1}{\left[ \sum_{j=1}^{4096}
    \left(\sum_{k=0}^3 z^{N_{j,k}}
    \omega_p^{P_{j,k}} \omega_b^{B_{j,k}}\right) \prod_{\ell=1}^9
    (R_\ell^M)^{E_{j,\ell}}\right]^4} \times 
  \frac{(g_{10}^{M+1}g_{11}^{M+1}g_{12}^{M+1})^2}
    {(g_{10}^{M}g_{11}^{M}g_{12}^{M})^{10}}.
\label{ry}
\end{equation}

Now we may use the recursion relations Eqs. (\ref{rr}) to express the
partial partition functions for 
subtrees with $M+1$ generations in terms of the ones with $M$
generations, and finally will arrive at the expression for the second
fraction in expression (\ref{ry})
\begin{equation}
\frac{(g_{10}^{M+1}g_{11}^{M+1}g_{12}^{M+1})^2}
    {(g_{10}^{M}g_{11}^{M}g_{12}^{M})^{10}}=\prod_{i=10}^{12}
\left[\sum_{j=1}^{1024} \left(\sum_{k=0}^3 z^{n_{j,k}}
\omega_p^{p_{i,j,k}} \omega_b^{b_{i,j,k}}\right) \prod_{\ell=1}^9
(R_\ell^M)^{e_{i,j,\ell}}\right]^2.
\end{equation}
\end{widetext}

Therefore, we see that we may express the bulk free energy per hexagon
as a function of the parameters of the model and the ratios $R_i$, and
in the thermodynamical limit it will converge to a fixed point
value. Finally, since we are in the grand-canonical ensemble, we have
that the pressure is $P=-\Phi/V$, where $\Phi$ is the grand-canonical
potential and $V$ the volume. Associating a volume $v_0$ to each site
of the lattice and recognizing $\phi_b$ as the grand-canonical
potential {\em per hexagon} for the solution on the Husimi tree, the
pressure may be written as
\begin{equation}
 P=-\phi_b/{4v_0},
\label{pres}
\end{equation}
where it should be stressed that we have four sites per hexagon in the
core of the tree.

\section{Thermodynamical behavior of the model}
\label{tp}
To study the thermodynamical behavior of the model on the Husimi
lattice, we define reduced intensive or fieldlike thermodynamic variables 
(temperature, pressure and chemical potential) as ${\bar
  T}=k_BT/\epsilon$, 
${\bar\mu}=\mu/\epsilon$ and ${\bar P}=Pv_0/\epsilon$. 	Considering expressions
(\ref{phib}) and~(\ref{pres}), the reduced pressure is given by:
\begin{equation}
{\bar P}={\bar T} \frac{\ln\left(\frac{Y_{M+1}}{Y_M^5}\right)}{8}.
\end{equation}

As mentioned before, we limited our study on the particular
case ${\bar \gamma}=2$, for which MC simulations were found in the
literature. For  
fixed values of ${\bar T}$ and ${\bar \mu}$, we iterate the recursion
relations~(\ref{rrr}) for the ratios of partial partition functions,
and once the 
fixed point is reached, we determine the mean numbers of particles, hydrogen
bonds and van der Waals interactions per lattice site, which are
represented by $\rho$, $\nu_{hb}$, and $\nu_{vW}$, respectively. The
densities of hydrogen bonds and van der Waals interactions per site,
normalized 
to be in the range $[0,1]$, will be $\rho_{HB}=\nu_{HB}/2$ and
$\rho_{vW}=\nu_{vW}/3$. Finally, the pressure may be also obtained at
the fixed point.

\begin{figure}
\begin{center}
\includegraphics[scale=0.62]{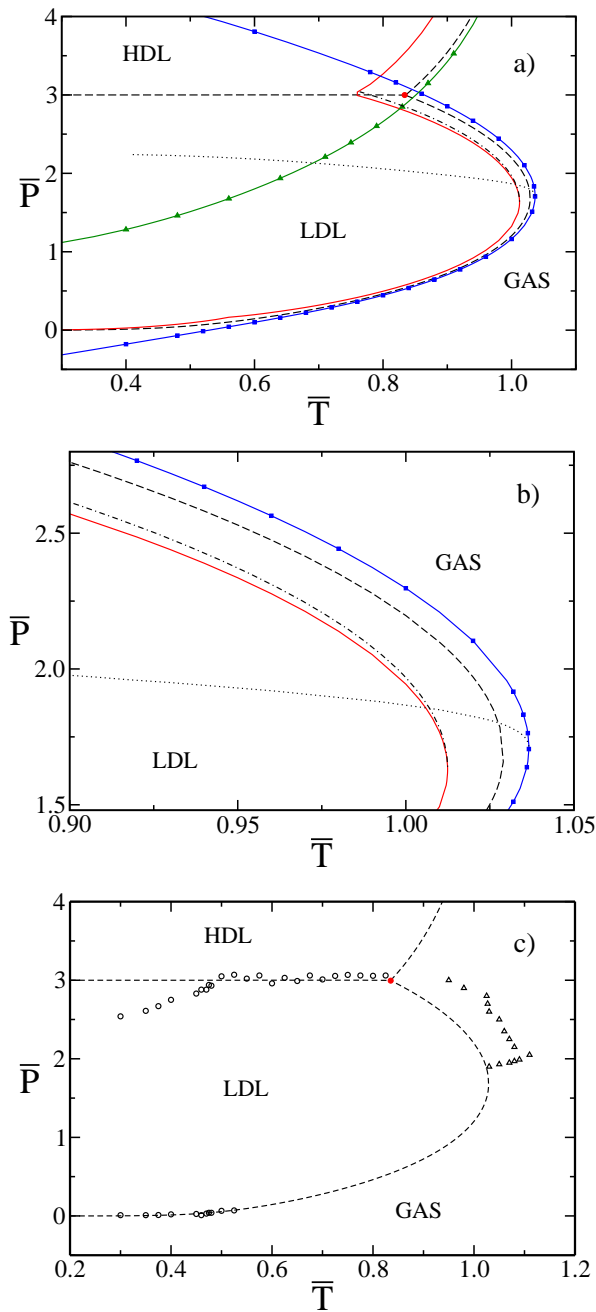}
\caption{(color on line) a) Phase diagram (${\bar T} \times {\bar P}$)
  of the Henriques-Barbosa model on the Husimi lattice, as defined on
  section~\ref{mod}. Dashed lines are discontinuous transitions which 
  meet at  triple point represented by a full circle (red
  on-line). The full lines are the stability limits of the GAS (no
  symbol, red on line),
  LDL (squares, blue on line), and HDL (triangles, green on line)
  phases. The dotted and dash-dotted lines (black on line)
  are the TmD and TMD, respectively. b) A detail of Fig. a) where the
  lines of density anomalies are more visible. 
  c) Present results for the coexistence lines and triple
    point are compared to the first order phase transitions (circles)
    and TMD (triangles) from Monte Carlo simulations of Balladares
    \textit{et al} \cite{b07}.} 
\label{f4}
\end{center}
\end{figure}

The phase diagram of the model was found using this procedure,
  being presented in the 
$({\bar T},{\bar P})$ plane on Fig. \ref{f4}. The three phases used in
our ground state analysis were also found at finite temperature and  
coexistence lines between these phases were calculated by requiring
the identity of their bulk free energies. 
 All transitions are discontinuous  and a triple point, located at
 ${\bar P}=2.997$, 
${\bar T}=0.835$, and ${\bar \mu}=1.959$, was found. The coexistence lines 
meeting at the triple point satisfy the thermodynamical requirements for
this situation, such as the 180 degree rule \cite{w74}.

Our phase diagrams are qualitatively different from the one
  originally suggested based on Monte Carlo simulations \cite{b07},
  where the two coexistence 
lines start at low temperatures and end at critical points. They are
also different from the cluster variational results obtained
  for several waterlike models on the $bcc$ lattice\cite{bs08,p09},
  where the coexistence lines end at tricritical 
points. In Fig.~\ref{f4}(c) the MC simulation
results presented in \cite{b07} are also shown and a good agreement on
the location of the phase transitions  
is found between those and our results, except on the low-temperature
region of the 
LDL-HDL coexistence line. Along this region the simulations present a rather
large positive slope, and since the ground-state value is exactly
known (${\bar P}=3$), a large negative slope should occur in the
LDL-HDL coexistence line at 
low temperatures. This effect is even enhanced if
we recall that, due to the third law of thermodynamics and the
Clausius-Clapeyron equation, the coexistence curve has to be
horizontal at vanishing temperature \cite{footn03}. A similar situation
was found in 
the simulations of the model with distinction between donor and
acceptor arms \cite{h05}, where this point is discussed, particularly
with respect to the implications of the Clausius-Clapeyron
relation. Although these simulations have been recently
  revisited \cite{s09}, the 
new results for the LDL-HDL coexistence line do not include temperatures low
enough to reach the region we are discussing here. In our 
calculation, the HDL-LDL coexistence curve starts with zero slope at
vanishing temperature. As the temperature increases, the slope
has a small positive value, then the curve presents a maximum and the
slope becomes negative close to the triple point. These features are
not visible in the scale of Fig. \ref{f4}. 
It is interesting to notice that the estimated 
location for the LDL-HDL critical point in the simulations is quite
close to the triple point in our solution.

We carefully verified if
the transitions are actually 
discontinuous by studying the stability limits of the fixed points
associated to each phase. These
limits may be found calculating the jacobian of the recursion
relations~(\ref{rrr})
\begin{equation}
J_{i,j}=\left( \frac{\partial R_i^{M+1}}{\partial R_j^M}\right),
\end{equation}
at the fixed point ($M \to \infty$), and then requiring the absolute
value of the largest eigenvalue of the jacobian to be equal to
one. In  Fig.~\ref{f4}(a) the stability limits of all 
phases are shown. Although in part of the GAS-LDL coexistence
line the stability limit of the LDL phase is very close to the transition,
they are never coincident, thus assuring the discontinuity of the 
transition.

In order to find out if the stability limits of the fixed points are
in fact the spinodals (thermodynamic stability limits), we also
calculated the eigenvalues of the hessian associated with the
phases. For the ordered phases (LDL and HDL) we found a good numerical
coincidence of the spinodals and the stability limits everywhere. For
the GAS phase, at low temperatures, we were able to assure numerically
the coincidence between these curves, but at higher temperatures we had
numerical problems to evaluate the elements of the hessian, which are
second derivatives of the potential. 

\begin{figure}
\begin{center}
\includegraphics[scale=0.5]{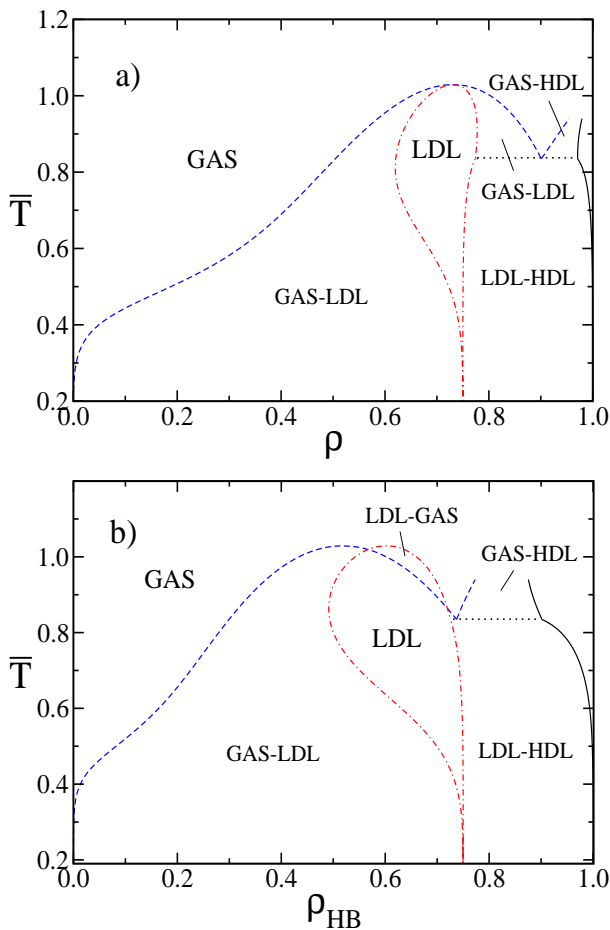}
\caption{(color on line) Temperature $\times$ density diagrams. a)
  Particle density. b) Hydrogen bond density. The temperature of the
  triple point is indicated by a dashed line. Densities of the phases
  at coexistence are indicated. The full line (black on line)
  corresponds to the HDL phase, the dashed line (blue on line) to the
  GAS phase and the dot-dashed line (red on line) the the LDL
  phase. The dotted horizontal line corresponds to the three-phase
  coexistence (triple point).
} 
\label{f5}
\end{center}
\end{figure}

\begin{figure*}
\begin{center}
\includegraphics[scale=0.63]{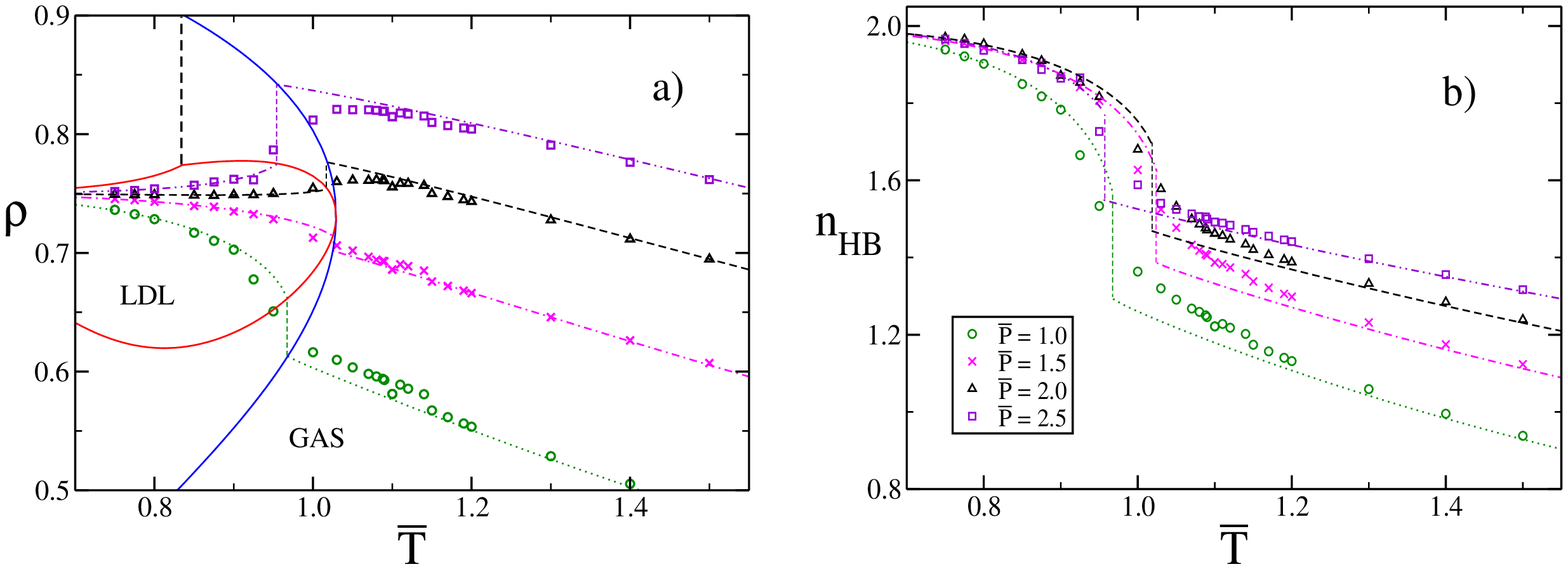}
\caption{(color on line) Isobaric curves of densities as a function of
  temperature. Broken lines are the results obtained in this work,
  with the vertical tielines also indicated at coexistence. Symbols
  are results from simulations by Balladares {\em et al} \cite{b07}}
\label{f6}
\end{center}
\end{figure*}

An interesting point is that
the LDL-GAS coexistence line has two regions with slopes of different
signs, showing a reentrant behavior. The ${\bar \mu}
\times {\bar T}$ phase diagram is quite similar to the ${\bar P}
\times {\bar T}$ phase 
diagram shown in Fig. \ref{f4}. The change of the sign happens at a
point which is located at ${\bar \mu}_{max}=0.137$,
${\bar T}_{max}=1.029$, and ${\bar P}_{max}=1.686$. 
Since the GAS phase has a larger entropy at the coexistence with the LDL 
phase, the Clausius-Clapeyron relation indicates that the particle
density should be 
lower for the GAS phase than for the LDL phase in the part of the
coexistence curve with pressures lower than ${\bar P}_{max}$.  
At $({\bar T}_{max},{\bar P}_{max})$, the densities of both
phases are identical and in the remainder of the coexistence the
density of the GAS phase is higher. In fact, the equal densities at
this point are confirmed in Fig. \ref{f5}(a), where  the temperature is
shown as a function of the density of particles at coexistence. It is
important to remind that GAS 
and LDL phases are not identical on this point (in this case it would
be a critical point). As can be observed in the phase diagram with the 
density of hydrogen bonds, instead of the particle density, shown in
Fig.~\ref{f5}(b). Although not presented here, densities of vdW
interactions are also different for both phases on this point.

Another relevant question is the location of the points of maximum
density in the isobars. We found that isobars for
the densities of particles as functions of the
temperature do actually present a maximum at pressures above ${\bar
  P}_{max}$. This TMD is located in the metastable extension of the
GAS phase inside the LDL phase, in Figs.~\ref{f4}(a) and~\ref{f4}(b), we 
represented the location of these metastable TMD of
the GAS with the dash dotted line. It ends at the point of maximum
temperature of the GAS 
  spinodal, as may be seen in the detail (Fig.~\ref{f4}(b)). We also
  found a temperature of 
{\em minimum} density line (TmD) 
  inside the LDL phase, shown in Figs.~\ref{f4}(a) and (b) as a dotted
  line. This line occurs also at pressures higher than ${\bar P}_{max}$ and,
unlike the TMD, 
which is located in the metastable GAS phase, the TmD line covers
both stable and metastable 
regions of the LDL, ending at the point of maximum temperature of the
LDL spinodal, a detail also more visible in Fig. \ref{f4}(b).  
It is actually expected that lines of vanishing thermal expansion
coefficient should end at the points where the corresponding spinodals
change the sign of their slope \cite{s82}.
Finally, it should be mentioned that similar findings 
were reported by Pretti and Buzzano in their homogeneous cluster
variational study of the symmetric Roberts-Debenedetti
model \cite{p04}.

In Fig.~\ref{f6} we show some isobars for the densities of particles
and hydrogen bonds. The results of the present calculations are
represented by the broken lines, and the symbols close to the
isobars are the MC simulation results obtained by
Balladares {\em et al} 
\cite{b07}. We notice a good quantitative agreement between them and our
calculations, at least not too close to the coexistence line. In the
${\bar P} \times {\bar T}$ diagram 
shown in Fig. 
\ref{f4}(c), the locations of the TMD points found in the
simulations presented in \cite{b07} are represented by triangles, and
in general we may notice that they are located at temperatures larger
than the ones of coexistence. 
These estimates actually correspond to the
maxima in the density at the coexistence curve, and the
fact that they lie above the coexistence curve may be due to
finite-size effects. As another possibility, the Bethe lattice
approximation introduced here could underestimate the location of the
TMD due to 
the absence of a LDL-GAS critical transition, observed in
simulations, at least for the model with distinction between donors
and acceptors \cite{s09}. In principle, the presence of such a
critical line could 
increase the entropy-volume cross fluctuations and shift the TMD line
($Vk_B T \alpha=\langle \delta V \delta S \rangle = 0$) to higher
temperatures. For example, this kind of TMD underestimation happened  
in Bethe lattice solution of the Bell-Lavis model of liquid
water \cite{b08}, when compared to Monte Carlo
simulations \cite{f09}.

In Fig.~\ref{f6}(b) the mean number of hydrogen
bonds per particle ($n_{HB}=\nu_{HB}/\rho=2\rho_{HB}/\rho$) is
depicted at constant pressure as a function of the temperature. Again, 
a good quantitative agreement between our results and the MC
simulations was found. In the simulations, crossing of different
isobars was 
reported and its physical origin was discussed \cite{b07}, in
relation to density anomaly. 
Nevertheless, our calculations suggest the possibility of the crossing being
a consequence of the discontinuous transition between the LDL and GAS phases and
finite-size rounding effects in the 
isobars.

\section{Final discussion and conclusion}
\label{conclu}
In this paper we solved the Henriques-Barbosa model with symmetric
arms \cite{b07} on a Husimi lattice built with hexagons. 
Two liquid and one gas phase are present in the phase diagram, but qualitative
differences are found when compared with the phase diagram which was
obtained with simulations. All transitions we found are discontinuous,
and also a triple point was found where all phases coexist with
different densities. We carefully checked if the transitions are
really discontinuous, since in part of the coexistence \textit{loci} the
discontinuities are rather small. Thus we assured that the stability
limits of the fixed points associated to the coexisting phases are
never coincident. Also, it may be seen in Figs. \ref{f5}
that the densities present a discontinuity at the coexistence line,
although it may be rather small, particularly in the
neighborhood of the point of maximum temperature in the GAS-LDL 
coexistence line. 

Recently, more detailed simulations were reported on this
model with distinction between donor and acceptor arms \cite{s09},
and a diagram closer to the one we present here was found. The
difference is that the GAS-LDL transition line is discontinuous at low
temperatures, but becomes a critical line when the temperature is
increased, thus a tricritical point is present. Also, the HDL-GAS 
transition appears to be continuous in the new simulations. The
LDL-HDL line is always discontinuous, so that in the simulations the
point which corresponds to the triple point in our phase diagrams
appears as a bicritical point (which was called wrongly as a
tricritical point in the caption of Fig. 3 in reference
\cite{s09}). Nevertheless, a direct comparison between the present
calculation and these new simulations may not be done, since the
distinction between donor and acceptor arms leads to an increase of
the entropy of the model. It is not impossible that
a transition found to be discontinuous in mean-field like approximations
turns out to be continuous in simulations or more precise
calculations, such as series expansions. However, we notice that the
results of the calculations presented here show, in general, 
good agreement with data furnished by simulations, as was also noticed
by Buzano and collaborators in their study of an associating
  lattice gas model, with 
tetragonal symmetry, on the $bcc$ lattice \cite{bs08}. It is worth
mentioning that the phase behavior presented by  
the most recent simulations of the Henriques-Barbosa model
is different from that found for the associating lattice gas model
studied with the cluster variational method on
Ref. \cite{bs08}. There,  
instead of a bicritical point, a tricritical and a critical endpoint
are present. We are presently studying this model
with the same methods implemented on this paper. 

Although the results presented here for the coexistence lines in the
pressure-temperature 
phase diagram agreed well with the data of simulations in
\cite{b07}, this is not the case for the low-temperature region of the
LDL-HDL coexistence curve, where the simulations suggest a 
minimum while a smooth behavior was found in our results. The
interpretation of this apparent minimum in relation with the
Clausius-Clapeyron equation seems unclear to us, and we believe on
the possibility that these results might be spurious. Possibly, longer 
equilibration times for MC simulations should be considered on this
low-temperature region.

If we adopt the qualitative phase diagram which emerges
from our calculations, the TMD found in the simulations would correspond to
the coexistence line. 
Nevertheless, it is also possible that the absence of a
critical line results in an overall decrease of the temperatures of
the TMD line, as in the case of the Bell-Lavis
model \cite{b08,f09}.
The crossing of isobaric curves for the density
of hydrogen bonds as function of the temperature was observed in the
simulations, but here it may be seen as a
consequence of rounding finite size effects for the discontinuous
transition at the LDL-GAS coexistence 
curve, with no relation to the TMD. The LDL-GAS coexistence curve
actually is a very weak discontinuous transition in the region where
the simulations suggested the presence of a critical point. This
indicates that the question of the order of the transition in this
region should be studied very carefully in simulations. 

The results
presented here support that the phase diagram of the symmetric
Henriques-Barbosa model does not present a second critical point, as
originally suggested through Monte Carlo simulations \cite{b07}. Our 
theoretical phase diagram is actually closer to the results of more
recent simulation of a 
original (asymetrical) model \cite{s09}: phase
transitions are found in both cases, but second order transitions are
not present in our theoretical results, this may be due to the
limitation of the Husimi lattice solution to capture long-range
correlations.

Considering the technical aspects of the Husimi lattice, our
calculation shows the 
importance of a careful choice of the sublattice structure to capture
the correct phase diagram of not so simple lattice models. A simpler
homogeneous lattice structure would lead the system to thermodynamic
states which may be unstable or metastable in the more general
parameter space used here (results not shown). On this sense, it is
worth mentioning that usually the simplest choice for the plaquette in
Husimi lattice calculations to approximate the behavior of models on
regular lattices is the elementary polygon in the original
lattice. Therefore, the simplest choice in the present case would be a
Husimi lattice with triangular plaquettes and coordination number
equal to 6. However, such a choice would probably not lead to results
comparable to the ones obtained in simulations. The rather unusual
choice of plaquettes we adopted here is certainly responsible for the
good quantitative agreement between our results and the
simulations. This technical discussion is extremely relevant to the
subject since one can find in the literature lattice models presenting
water like behavior (including thermodynamic anomalies and
interesting phase diagrams) which were studied without a more detailed
sublattice analysis \cite{f02,c08}. 

We finish remarking that many lattice models proposed to
  investigate waterlike anomalous behavior were found  
to present phase diagrams which were more complex than originally
expected. When sublattice ordering is properly considered \cite{p09},
even the simplest  
models do present at least a single critical line, and many of them do
not present gas-liquid phase transition ending in a critical point  
or a temperature of maximum density in a stable disordered fluid
without a sublattice structure. The arbitrary use of the   
homogeneneity assumption can result in some interesting phase diagrams
but, in our opinion, this assumption is an artifact which eventually
may hide the instability of the homogeneous phase at low temperatures.
 Thus, this homogeneous solution could not be considered as    
a true implementation for a random lattice useful for representing
fluids, as recently proposed \cite{p09}.  
Considering the results obtained here and in other recent
papers \cite{bs08,p09}, the issue of finding a simple lattice model  
with minimal waterlike behavior seems to be far from being
resolved. Monte Carlo simulations are certainly needed to determinate  
the `exact' phase diagram but it seems that a modeling breakthrough
will be needed to achieve a better description of liquid water. 

\section*{Acknowledgements}
TJO acknowledges doctoral grants by CNPq and FAPERJ. JFS is grateful
to CNPq for partial financial support and MAAB acknowledges financial
support from FAPESP and CNPq. We thank Profs. M\'{a}rcia 
C. Barbosa and Vera B. Henriques for helpful discussions. We also
thank Prof. M\'{a}rcia C. Barbosa for providing us MC simulation
data.


\begin{thebibliography}{99}
\bibitem{footn01}A typo in the expression (2) of reference \cite{b07}
  should be corrected: the factor $u$ before the second sum should be
  replaced by $-2u$.
\bibitem{footn03}This is not the case of
  systems which do present residual entropy. As an example, the model
  studied in ref. \cite{b08} displays a coexistence line with finite
  slope because the high density liquid has a residual entropy, whose
  origin is related to presence of frustration.
\end{thebibliography}
\end{document}